\documentclass[preprint,12pt]{elsarticle}

\usepackage{graphics}

\usepackage{epsfig}

\usepackage{amssymb,amsthm}

\journal{arXiv}

\begin{document}

\begin{frontmatter}

\title{Eye-Safe Solid-State Quasi-CW Raman Laser with Millisecond Pulse Duration}

\author[labelAff]{Anton~A.~Kananovich\corref{cor1}}
\ead{kananovich@tut.by}
\author[labelAff]{Vladimir~I.~Dashkevich}
\author[labelAff]{Pavel~A.~Apanasevich}
\author[labelAff]{Dmitriy~H.~Zusin}
\author[labelAff]{and~Valentin~A.~Orlovich}

\address[labelAff]{B.I.~Stepanov Institute of Physics, National Academy of Sciences of Belarus, 220072, 68 Nezalezhnasti av. Minsk, Belarus}

\cortext[cor1]{Corresponding author}

\begin{abstract}
We demonstrate the first quasi-CW (ms-long pulses, pump duty cycle of 10\%) end-diode pumped solid state laser generating eye-safe radiation via intracavity Raman conversion. The output power at the first Stokes wavelength (1524~nm) was 250~mW. A theoretical model was applied to analyze the laser system and provide routes for optimization. The possibility of true CW operation was discussed.

\end{abstract}


\end{frontmatter}


\section{Introduction}
\label{lIntro}
In recent years, much attention has been paid to solid state Raman lasers \cite{cerny2004solid,pask2003design,pask2008wavelength}. Demonstration of continuous-wave (CW) generation in such lasers is a technically hard task because high-power sources of CW laser radiation are needed to reach the threshold of stimulated Raman scattering (SRS). Additional requirements for the sources are narrow linewidth and low divergence of generated radiation. The first CW solid-state Raman laser was reported relatively recently \cite{grabtchikov2004multimode}. A high-power argon-ion laser was used as a pump source, a barium nitrate crystal placed in a separate high-Q cavity was used as a Raman-active medium. The next step in the development of Raman-based solid-state compact CW laser systems was the demonstration of intracavity Raman conversion in end-diode-pumped lasers \cite{demidovich2005continuous}. To date, in this approach two effects were implemented: self-frequency Raman conversion \cite{demidovich2005continuous} and Raman frequency conversion \cite{pask2005continuous}. In a self-frequency Raman laser, fundamental generation and Stokes shift take place in the same crystal. In case of Raman frequency conversion, generation of fundamental radiation and Stokes shift take place in two different media. Also, self-frequency Raman conversion in a composite crystal was realized \cite{lisinetskii2007NdKGW}. Up to 3.4~W of the output power at the first Stokes wavelength was obtained \cite{fan2009high}. For generation of the fundamental radiation, all of the published CW diode pumped solid state lasers with intracavity Raman conversion (CWSSRL) made use of the strong $^4F_{3/2} \rightarrow ^4\!\!I_{11/2}$ transition of $\rm Nd^{3+}$ ions resulting in radiation at the wavelength of about $1.06\mbox{ }\mu$m. Using different combinations of laser and Raman-active crystals one can obtain generation at different discrete wavelengths near 1.2~$\mu\mbox{m}$. In such lasers, it is also possible to obtain visible radiation within 560-590~nm spectral range by intracavity second harmonic generation of the Stokes radiation \cite{dekker2007all} and sum-frequency generation of the Stokes and fundamental radiation \cite{kananovich2009generation,kananovich2010all}. Recent demonstration of second Stokes generation in a CWSSRL \cite{lee2010near} preceded by the first demonstration of second Stokes in an externally pumped CW Raman laser \cite{grabchikov2009continuous} has further increased the number of wavelengths obtainable with such lasers and, subsequently, expanded the range of CWSSRL applications.

In this paper, we demonstrate what we believe is the first quasi-CW (ms-long pulses) end-diode-pumped solid state laser with intracavity Raman conversion in which fundamental radiation is generated on $^4F_{3/2} \rightarrow ^4\!\!I_{13/2}$ transition of $\rm Nd^{3+}$ ions (1.5~$\mu$m CWSSRL) as opposed to $^4F_{3/2} \rightarrow ^4\!\!I_{11/2}$ transition in conventional CWSSRL (1.2~$\mu$m CWSSRL) reported so far. Being in the so-called eye-safe wavelength range, the resulting wavelength of 1524~nm is applicable in remote sensing, optical fiber communication, medicine, meteorology, range finding, and etc. In this first demonstration of a 1.5~$\mu$m CWSSRL, thermal issues have not been fully resolved. To mitigate thermal effects we used quasi-CW (qCW) pumping. We also used double-end-pumped geometry with two crystals. The use of double-end-pumping effectively divides thermal load between two ends of a laser crystal (or two laser crystals) reducing the risk of fracture and improving power-scalability \cite{ogilvy2003efficient}. The model of a CWSSRL developed in \cite{apanasevich2010power} is applied to analyze the demonstrated laser. The ways to improve output characteristics as well as the issues of a true CW operation are discussed.


\section{Experimental}
\label{lExp}

\begin{figure}[ht]
\centering
\includegraphics[width=0.5\textwidth]{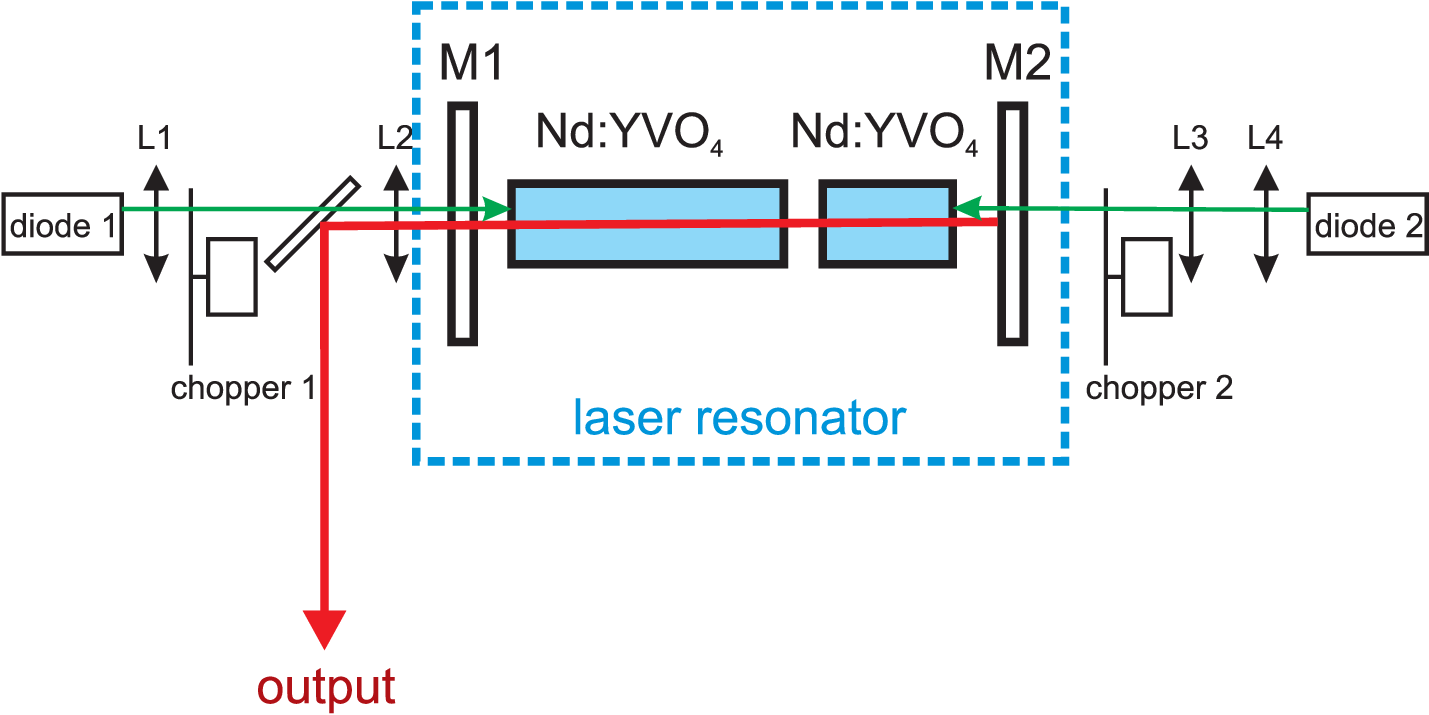}
\caption{Experimental setup}
\label{figSetup}
\end{figure}

The experimental setup is presented in Fig.~\ref{figSetup}. The output from the diode lasers (diode~1 and diode~2) was re-imaged into the laser crystals using collimating lenses L1 and L4, and focusing lenses L2 and L3. The resulting pump spot diameters in the laser crystals averaged over absorption length were $\approx 270\mbox{ }\mu\mbox{m}$ for both crystals. Chopper~1 and chopper~2 were inserted into the pump beam path. The choppers were synchronized to shut the beams of diode~1 and diode~2 simultaneously. With the choppers on, the pump pulse duration was 7.1~ms (FWHM) and the repetition rate was 14~Hz (duty cycle 10\%: 7.1~ms on, 63.9~ms off). Hereafter by pump power we mean instant power (on-power) incident on both crystals if not mentioned elsewise. 

The laser crystals were \textit{a}-cut $\mbox{Nd:YVO}_4$ (0.4 at. \% doping both). The longer crystal pumped by diode~1 had $5\times5\times20\mbox{~mm}^3$ dimensions. The shorter one was $3\times3\times12\mbox{~mm}^3$. All the crystal faces were antireflection coated for a broad range of 0.8-1.5~$\mu\mbox{m}$. The residual reflections of the coatings at 0.8-1.5~$\mu\mbox{m}$ were 0.1\%-0.5\% per each coating. The crystals were positioned so that their \textit{c}-axes were parallel and were placed in water-cooled copper blocks. The laser resonator consisted of flat identically coated mirrors M1 and M2. The mirrors had high transmission at the pump wavelength of 809~nm ($T_1^p = 95$\%) and 1064~nm to suppress oscillation at the strong $^4F_{3/2} \rightarrow ^4\!\!I_{11/2}$ transition. Both mirrors had high reflection at the laser wavelength of 1342~nm and the wavelength of the first Stokes line, 1524~nm ($R > 99.9$\% at both wavelengths). All the  elements inside the cavity were placed as close as possible; the overall geometrical length of the cavity was approximately 43~mm.

To deliver the output beam to recording systems a turning mirror was placed between the lenses L1 and L2. It was HR-coated at 1300-1550 nm ($R>97$\%) and HT-coated at 809~nm ($R<5$\%). The output power was measured using a power meter (Ophir LaserStar) with a thermoelectric head. Pulse shapes were recorded with a two channel Tektronix 3052B (500 MHz) oscilloscope equipped with fast InGaAs p-i-n  photodiodes. The spectra of the output radiation were recorded using a calibrated MDR~23 grating monochromator with Hamamatsu G9212-512Q CCD array (spectral sensivity 0.9-1.7~$\mu$m). A BeamOn IR1550 beam profiler (Duma Optronics LTD.) was used to measure spatial properties of the beams.

The threshold of Stokes generation was reached at 13.8~W of pump power. At the highest instant pump power of 56.8~W, the maximum output power reached 127 mW. Because both of the laser mirrors had identical spectral properties, we assume that similar powers were generated in the opposite directions and, thus, the overall output power at the Stokes wavelength reached 250~mW. The output power and its maximization is discussed in Section~\ref{lTheor}.

\begin{figure}[ht]
\centering
\includegraphics[width=0.5\textwidth]{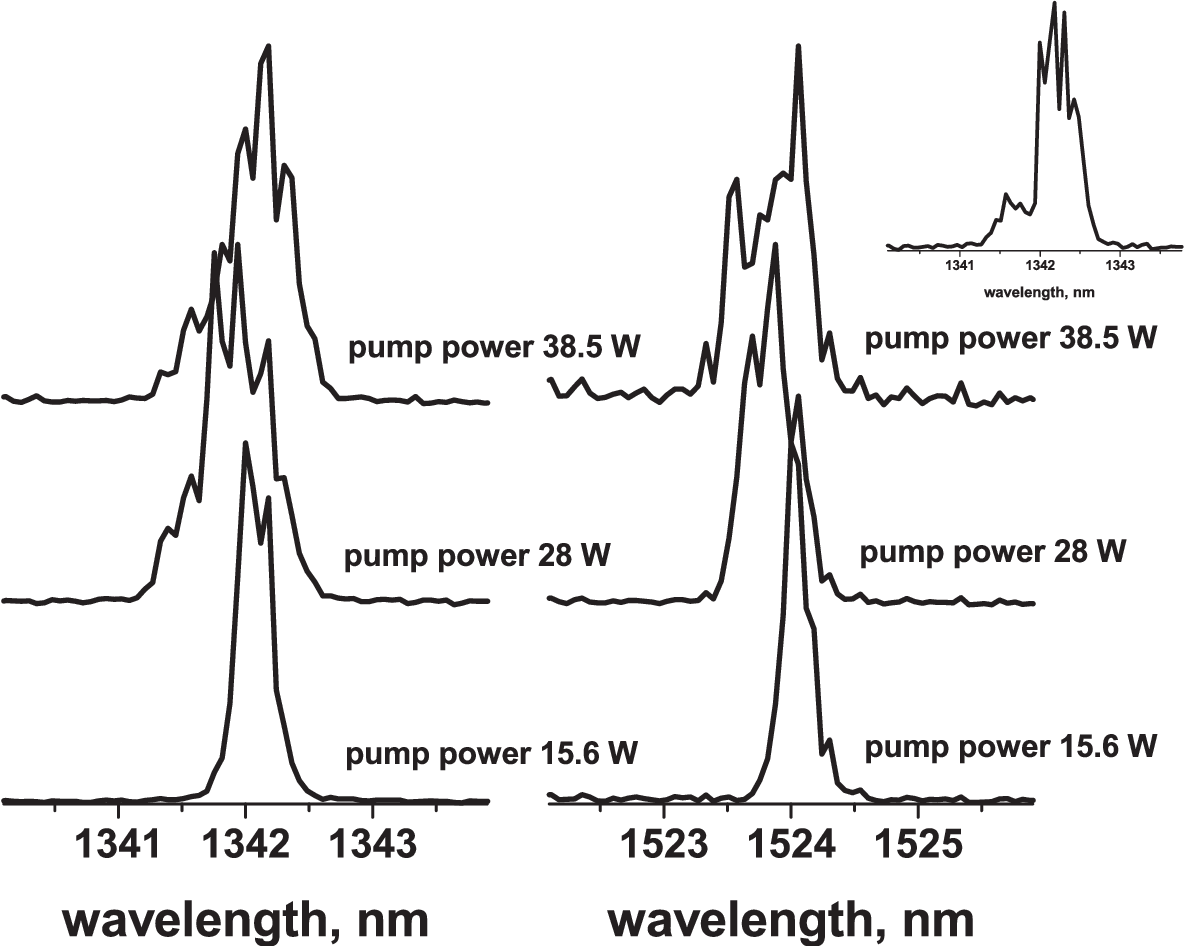}
\caption{Output spectra of the laser at different pump powers. Inset, the spectrum at the output power of 2 W (no SRS)}
\label{figSpe}
\end{figure}

The samples of the output laser radiation spectra at different diode pump powers are presented in Fig.~\ref{figSpe}. Both lines, the fundamental and the Stokes one, had FWHM of about 0.5~nm. Several subpeaks of the Stokes line were observed at pump power levels greater than 28~W. The fundamental line demonstrated subpeaks at any pump power available. We believe, the subpeaks are due to manifestation of the resonator axial modes. The separation of longitudinal modes in our resonator is about 0.011~nm for the fundamental line and 0.014~nm for the Stokes line. We observed 2-4 subpeaks with the separation of 0.12-0.42~nm. It should be noted that the resolution of our spectrometer was only 0.06~nm. In paper \cite{dekker2007continuous}, the spectra of 1.2~$\mu$m CWSSRL were recorded with much higher resolution and many more peaks were visible for both the fundamental and the Stokes lines. For the laser discussed in the present paper the situation must be the same. Positions of the subpeaks observed by us changed from one realization to another (the exposure of spectrum recording was 300~ms, i.e. an integral spectrum for 50 pulses was recorded). This means that the process of qCW generation of the fundamental radiation with high mirror reflections at 1342~nm is accompanied by changes in longitudinal resonator modes composition, i.e. each time different groups of modes lase. The spectral distribution of the Stokes lines had fewer peaks because Raman conversion was efficient only for the most intense spectral parts of laser radiation.

\begin{figure}[ht]
\centering
\includegraphics[width=0.5\textwidth]{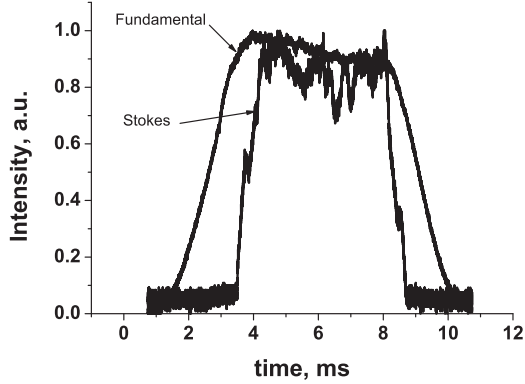}
\caption{Oscilloscope traces}
\label{figOsc}
\end{figure}

The Stokes generation was stable at all pump powers exceeding the threshold. An example of Stokes and fundamental waveforms at 28~W pump power is shown in Fig.~\ref{figOsc}. One can see that the oscillation of the eye-safe Stokes radiation lasted for $\sim$5~ms within each pump pulse. The Stokes pulses are characterized by rather fast pulse edges compared to the laser ones. The Stokes field lases almost immediately when the fundamental intensity reaches its maximal value. A delay between the starts of the laser and Stokes generation, predicted in paper~\cite{grasjuk1974generationEn} and observed in experiment for the first time in~\cite{demidovich2005continuous}, is not seen in the Fig.~\ref{figOsc} as it is overshadowed by the slow build up of the fundamental field owing to effective slow turn on of the pump power due to the slow mechanical chopper. In Fig.~\ref{figOsc} one can see that the Stokes oscillation appears to be noisier the fundamental one. That could be explained by the fact that the laser operated not far from the threshold of the SRS and the Stokes field didn't deplete the fundamental. Thus small fluctuations in the fundamental intensity lead to the large ones in the Stokes whereas fluctuations of the Stokes field didn't affect much the fundamental one.

At high pump powers the quality of the fundamental beam was deteriorated, the laser apparently oscillating at several transverse modes, while the Stokes beam preserved better quality (beam-cleanup effect). The $\rm M^2$ parameters measured at 28~W of pump power were 2.7 and 6.5 for the Stokes and fundamental beams respectively. It should be noted that in the absence of Raman conversion (near the Raman oscillation threshold) the beam quality factor of the fundamental beam was $\rm M^2_{L} =4.9$. With a small excess over the threshold the parameter $\rm M^2_{L}$ was 5.6.

\section{Modelling and discussion}
\label{lTheor}

We used the model developed in \cite{apanasevich2010power} to analyze the obtained experimental results and to discuss the possibility of true CW generation at 1524~nm in a CWSSRL. The model was developed under the following assumptions. The spectral output of the pump diode laser coincides with the laser medium absorption band. The pump beam is essentially fully absorbed at a single pass through the laser medium. The intracavity fundamental and Stokes beams are assumed to be coaxial with the pump beam. The frequencies of generated radiations are determined by operating transitions for the lasing and SRS processes (no detuning). For both generated lines the spectral width is assumed to be considerably smaller than that of the corresponding operating transition. Under these assumptions, the threshold pump power for generating Stokes radiation, as given by \cite{apanasevich2010power}, is: 

\begin{equation}
\label{eqThresh}
\mathop {\left( {P_p^{in} } \right)}\nolimits_{th}^s = \frac{\gamma _L 
}{2k_g \alpha _p A\beta T_1^p }\frac{\alpha _L C_{LL} \gamma _s }{2G\mathop 
{\overline M }\nolimits_{Ls} },
\end{equation}
where $\gamma _{L(s)}=- \frac{1}{2}\ln\left(R_{1}^{L(s)} R_{2}^{L(s)} \right)  + \gamma_{{int}_{L(S)}}$ is the overall loss at the fundamental (Stokes) wavelength, including transmissions of the input mirror $T_{1}^{L(s)} = 1- R_{1}^{L(s)}$ and the output mirror $T_{2}^{L(s)} = 1 - R_{2}^{L(s)}$, and overall internal losses at the wavelength of the fundamental (Stokes) radiation $\gamma_{{int}_{L(S)}}$, including the losses due to absorption and scattering in the active medium, diffractive losses and losses on the AR coatings; $l$ is the length of the active medium; $\alpha _{L}=\sigma _{L}T_{1}/h\nu _{L}$, $\sigma _{L}$ is the gain cross section at the center of the laser transition line, $T_{1}$ is the lifetime of the upper level of the lasing transition, $h$ is the Planck constant, $\nu _{L(s)}$ is the frequency of the fundamental (Stokes) radiation; $k_{g}=N\sigma _{L}$, $N$ is the density of active centers, $\alpha _p = (\sigma _p T_1 / h\nu _p )[\overline {\Delta \nu _p } / 
(\overline {\Delta \nu _p } + \Delta \nu _p )]$, with $\Delta\nu _p$ being pump spectral width and $\overline{ \Delta \nu _p }$ being spectral width of absorption band; $T_1^p$ is the transmission coefficient of the input mirror at the pump frequency; $\beta$ is the fraction of absorbed pump power that goes for the occupation of the upper laser level; and $G$ is the Raman gain coefficient. Parameters $A$, $\mathop{\overline M }\nolimits_{Ls}$, and $C_{LL}$ characterize overlap of the pump, laser, and fundamental beams. Under the assumptions stated above, for Gaussian beams the parameters take the form \cite{apanasevich2010power}:

\begin{equation}
\label{eqMLS}
\mathop{\overline M }\nolimits_{Ls} = \frac{l}{\pi  \left(w_L^2+w_S^2\right)},
\end{equation}

\begin{equation}
\label{eqCLL}
C_{LL} = \frac{w_L^2+w_P^2}{2 \pi  w_L^2 w_P^2+\pi  w_L^4},
\end{equation}

\begin{equation}
\label{eqA}
A = \frac{1- e^{-kl_{cr}}}{\pi k \left(w_L^2 +  w_P^2 \right) } ,
\end{equation}
where $k = N \sigma_p[\overline {\Delta \nu _p } / 
(\overline {\Delta \nu _p } + \Delta \nu _p )]$; $w_P$, $w_L$ and $w_s$ are the radii of the pump, fundamental and Stokes beams correspondingly.

According to the model, the output Stokes power is:

\begin{equation}
\label{eqPow}
P_{s2} = \eta [P_p^{in} - \mathop {\left( {P_p^{in} } \right)}\nolimits_{th}^s ],
\end{equation}
where $\eta = \frac{\nu _s }{\nu _L }\frac{k_g \alpha _p AT_2^s T_1^p \beta}{\alpha _L C_{LL} \gamma _s }$, and $P_p^{in}$ is the diode pump power.

We used (\ref{eqPow}) to estimate the Stokes output power of the laser under consideration as a function of the diode pump power.  The values of the parameters applied for the estimation are given in Table~\ref{tParam}.  All of the numerical values provided in the table are either known from the literature or measured. The exception is the intracavity losses $\gamma_{{int}_L}$ and $\gamma{{int}_S}$, which are very hard to measure in the case of CWSSRL. The value of the losses provided in the Table~\ref{tParam} were chosen to fit the experimental data. The numerical value of $w_P$ is the pump beam radius averaged over absorption length in the crystals. Radii $w_L$ and $w_S$ were found from cavity modeling by the ABCD method. To perform such modelling, one needs to know the focal length of the thermal lens in the active media. We can estimate the thermal lens focal length induced by the pump beam using the formula \cite{innocenzi1990thermal,chen1997optimization}:

\begin{equation}  
\label{eqInno}
\frac{1}{f}= \left( \frac{dn}{dT} + n \alpha_t \left( 1 + \nu \right) \right) \frac{1}{K_c} \frac{P_{abs}}{\pi (w_P)^2} \left( 1 - \frac{\lambda_p}{\lambda_{L}}  \right),
\end{equation}
in which the overall thermal response of $\mbox{YVO}_4$ is taken into account. In (\ref{eqInno}) $K_c = 5.23 \mbox{ W/mK}$ \cite{bermudez2002thermo} is the thermal conductivity of $\mbox{YVO}_4$, $\left( dn / dT \right) = 4.7 \cdot 10^{-6} \mbox{ K}^{-1}$ \cite{yao2005double} is the  thermo-optical coefficient, $n = 2.16$ \cite{weber2003handbook} is the refractive index, $\alpha_t = 4.4 \cdot 10^{-6}$~K${}^{-1}$ \cite{weber2003handbook} is the thermal expansion coefficient, $\nu = 0.3$ \cite{bermudez2002thermo} is the Poisson ratio, and $P_{abs}$ is the pump power absorbed in the active crystal. Formula (\ref{eqInno}) provides estimation of a thermal lens corresponding to the quantum defect between the pump and the laser. The inelastic SRS process provides additional heat generation. In some works (see \cite{dekker2007continuous,omatsu2009passively,lee2010efficient}) the total heat load is taken into account by replacing the wavelength $\lambda_{L}$ with the Stokes wavelength $\lambda_{s}$ in (\ref{eqInno}). In our view, this is not entirely correct because in SRS the portion of the heat generated in the active crystal is proportional to $1 -  \lambda_{L}/\lambda_{s}$. However, to simplify evaluation of $f$ for numerical analysis we also use the method of replacing  $\lambda_{L}$ with $\lambda_{s}$.

\begin{table*}
	\caption{Parameter values used in calculations}
	\centering
		\begin{tabular}{ c | c || c | c | }
			Parameter & Value & Parameter & Value \\ \hline
			$T_{1}^{L}$ & 0.1 \%                           & $w_P$  & 270 $\rm \mu$m \\
			$T_{1}^{s}$ & 0.05 \% 													 & $w_L$  & 150 $\rm \mu$m \\
			$T_{2}^{L}$ & 0.1 \% 													 & $w_s$  & 130 $\rm \mu$m \\
			$T_{2}^{s}$ & 0.05 \% 													 & $\sigma_p$ & $2.7 \cdot 10^{-19} \mbox{ cm}^2$ \cite{zagumennyi1996crystals}\\
			$l        $ & 32~mm  													 & $T_1^p$ & 95 \% \\
			$\gamma_{{int}_{L}}$ & 3 \%                            & $\gamma_{{int}_{S}}$  & 3 \%  \\
			$N        $ & $0.5 \cdot 10^{20}\mbox{ cm}^{-3}$  & $\lambda_p = \frac{c}{\nu_p}$ & 809~nm \\
			$\Delta\nu _p$ & $50 \mbox{ cm}^{-1}$  & $\overline {\Delta \nu _p }$ & $60 \mbox{ cm}^{-1}$ \cite{lieto2003high} \\
			G & $\rm 4.5 \frac{cm}{GW}$ \cite{kaminskii2001tetragonal}  & $\beta$ & 0.7 \\ 
		\end{tabular}	
	\label{tParam}
\end{table*}

To estimate the effective value of $P_{abs}$ in (\ref{eqInno}) at quasi-CW operation we applied approximations used in \cite{boyd2003nonlinear}. With these approximations, the heat transfer equation writes down as follows:

\begin{equation}
\label{eqBoyd}
\rho C\frac{{\partial T}}{{\partial t}} + {K_c}\frac{T}{{{w_P}^2}} = k I(t),
\end{equation}
where ($K_c$, $k$ and $w_P$ are given earlier in the text) $C=0.56$~J/gK \cite{sato2006studies} is the specific heat and $\rho = 4.23\mbox{ g/cm}^3$ \cite{weber2003handbook} is the mass density of the crystal, and $I(t)$ is the pump intensity representing a sequence of rectangular pulses with 71~ms period and 10\% duty cycle. In such a configuration with the incident pump power of 30~W, effective value of $P_{abs}$ was estimated to be 3.4~W. The estimation was made by relating the average temperature obtained from the solution of equation (\ref{eqBoyd}) to the maximum temperature given by the formula $T_{max}= k I_{max} w_{P}^2 /K_c$ \cite{boyd2003nonlinear}. The value $P_{abs} = 3.4$~W corresponds to thermal lens focal length $f=50$~mm, as calculated by (\ref{eqInno}). This value is almost 5~times higher than the length of the shorter crystal and 2.5 times higher than that of the longer one. Thus, the thin lens approximation implied by equation (\ref{eqInno}) is valid for case of qCW operation.  In the case of true CW one would probably need more complex calculations to estimate distributed thermal lens and its influence on the parameters of the beams inside the cavity.
 
\begin{figure}[ht]
\centering
\includegraphics[width=0.5\textwidth]{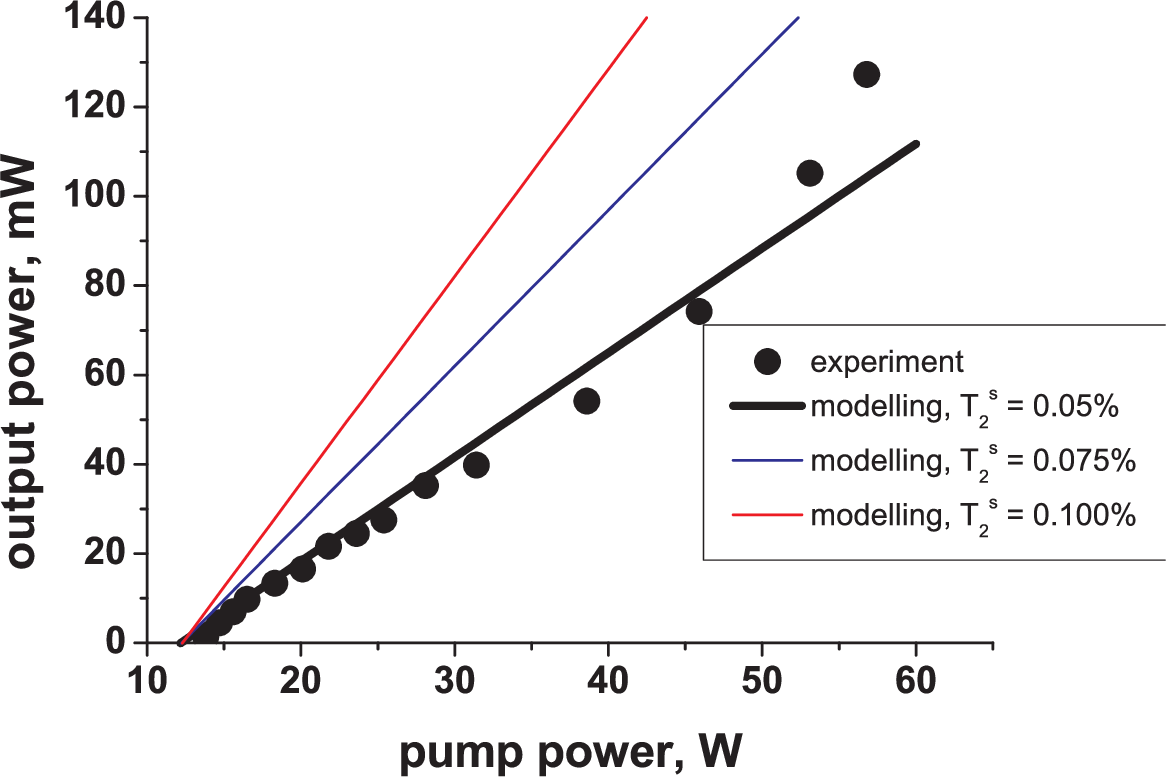}
\caption{Stokes output power as a function of the diode pump power. Dots represent the experimental data; lines is the estimation by the model.}
\label{figPow}
\end{figure}

The results of calculations of the Stokes output power  as a function of the diode pump power as well as the experimental data are presented in the Fig.~\ref{figPow}. To simplify calculations we used fixed values of the beam sizes ($P_{abs} = 3.4$~W); thus the slope of the theoretical curve is fixed. (Variation of the beams overlap with the pump power in the model used is given in  paper \cite{apanasevich2011stationary}). Nevertheless, even with constant beam sizes the results provided by the model are in good agreement with the experimental data. The disagreement between the theoretical and experimental curves becomes noticeable only at high diode pump power. One can observe that the slope of the experimental curve apparently increases with pump power after $P_p^{in}$ exceeds 45~W. This is mainly due to the following reasons. First, the wavelength of diode lasers increases with the power increase and reaches 809~nm (absorption peak for $\rm Nd:YVO_4$ crystals) at high powers. Second, the variation of pump power results in the variation of thermal lenses in the active media leading thus to a change of beam sizes and, subsequently, to a change of the $\mathop{\overline M }\nolimits_{Ls}$, and $C_{LL}$ values.

\begin{figure}[ht]
\centering
\includegraphics[width=0.5\textwidth]{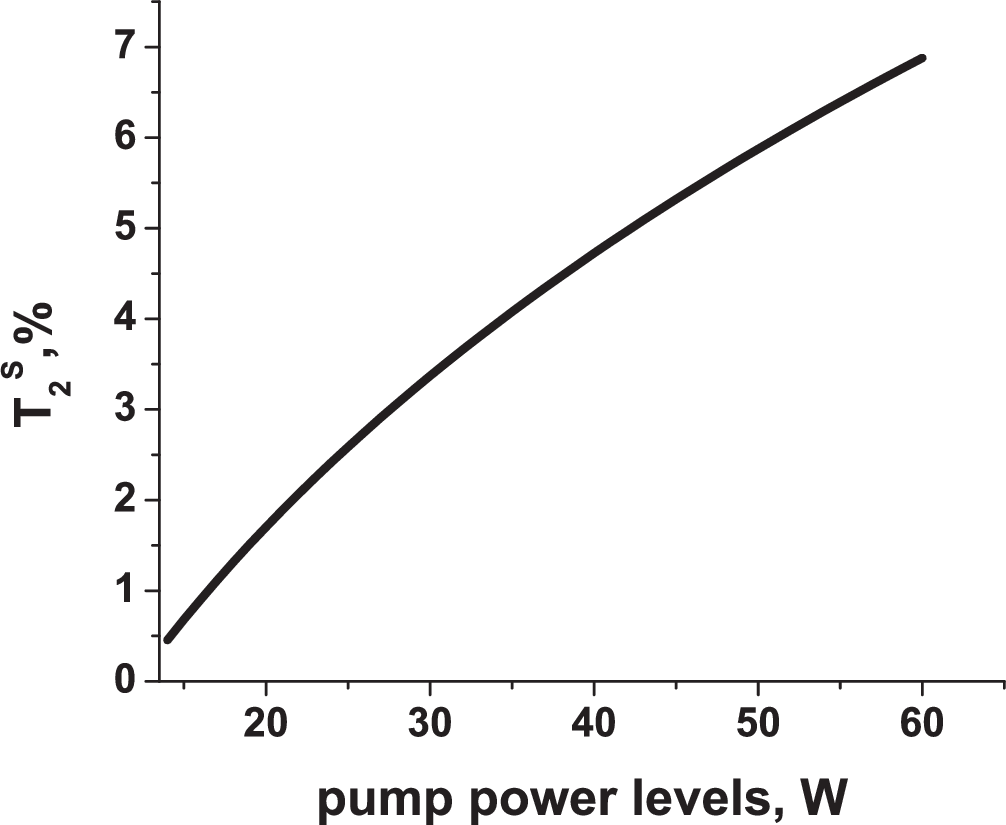}
\caption{Estimated output coupler optimal transmission at the Stokes wavelength.}
\label{figOptT}
\end{figure}

It is evident from Fig.~\ref{figPow} (color lines) that the Stokes output power of the present setup can be maximized by using output mirrors with higher transmission at the Stokes wavelength, yet this has to be verified in subsequent experiments. Large transmissions at the Stokes wavelength increase both slope and threshold; there exists an optimal transmission at every given level of pump power. Maximizing $P_{s2}$ in (\ref{eqPow}) with respect to $T_2^s$ one can calculate optimal transmissions of the output coupler for a given pump power level. The results of of such calculations for parameters of the current laser setup (see Table~\ref{tParam}) are presented in Fig.~\ref{figOptT}.

\begin{figure}[ht]
\centering
\includegraphics[width=0.5\textwidth]{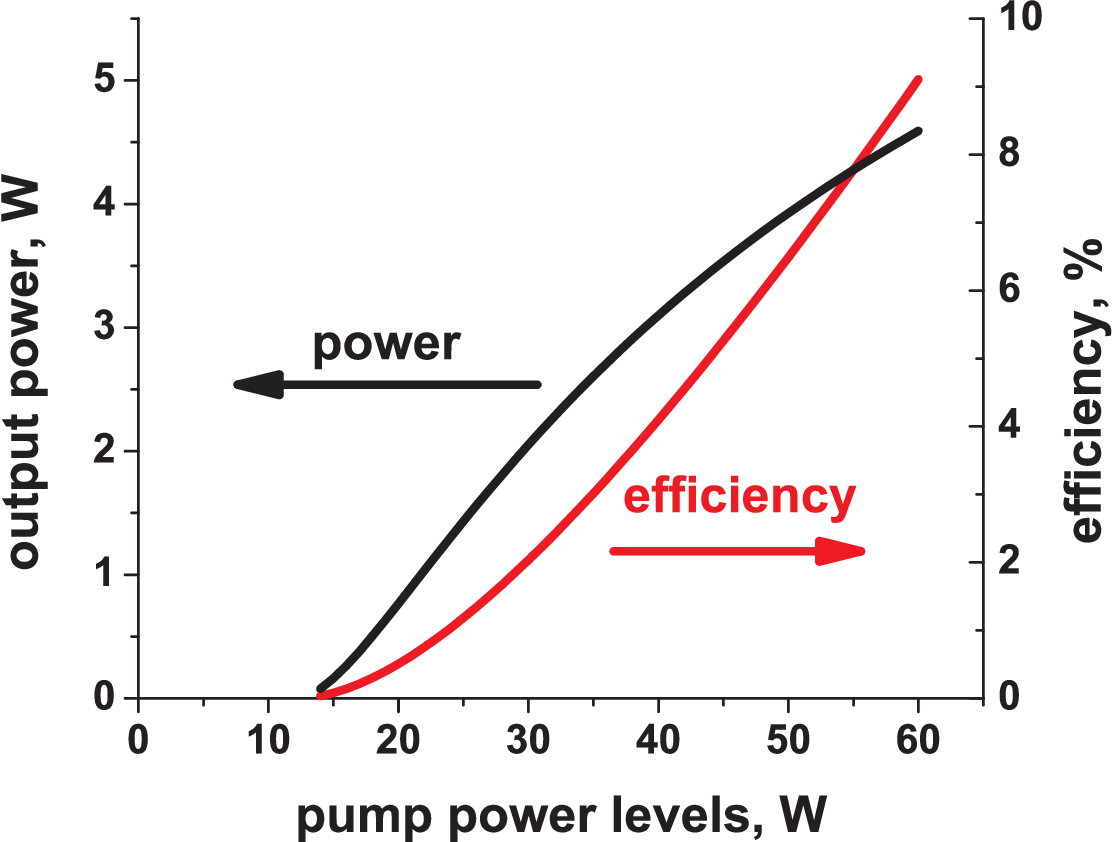}
\caption{Powers and efficiencies for optimal transmissions of the output coupler.}
\label{figMaxPowMaxEff}
\end{figure}

For example, for pump power level of 17~W the optimal transmission $T_2^s$ is 1.1\%, which is close to the one for a qCW 1.2~$\mu$m CWSSRL at the diode pump power level of the same order \cite{apanasevich2011stationary}. With an output coupler with this transmission the laser under consideration should provide 117~mW of Stokes output power at the diode pump power level of 17~W which corresponds to efficiency of 0.7\% with respect to the diode pump power. The maximum available diode pump power in the present setup was 57~W. For this pump power level the optimal transmission is 6.6\%, which according to (\ref{eqPow}) corresponds to the output power of  as much as 4.6~W and efficiency of 8\%. For other diode pump power levels, the output power and efficiency calculated with (\ref{eqPow}) using the optimal transmissions $T_2^s$ (indicated in Fig.~\ref{figOptT}) are presented in Fig.~\ref{figMaxPowMaxEff}.

\begin{figure}[ht]
\centering
\includegraphics[width=0.5\textwidth]{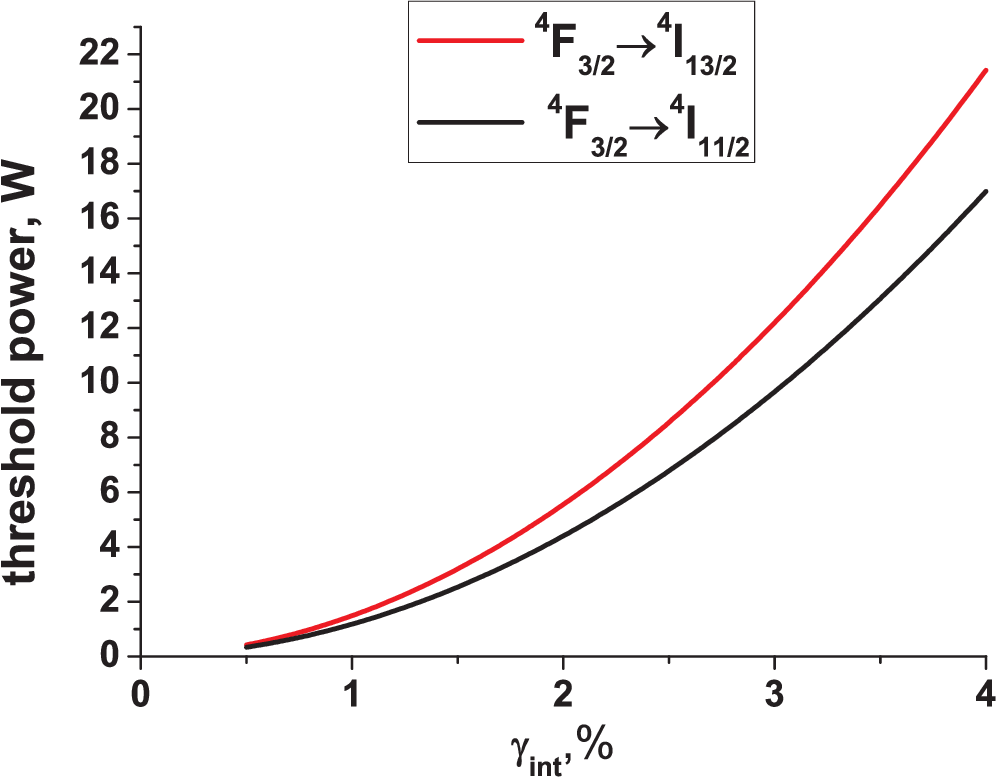}
\caption{The threshold of Stokes generation as a function of the intracavity losses $\gamma_{int}$.}
\label{figThresh}
\end{figure}

In experiment, we couldn't obtain Stokes generation in true CW regime. This could be explained by much higher diffractive losses in CW regime as compared to qCW.  The diffractive losses, which contribute to overall internal losses, are induced by highly aberrated thermal lenses (see for example \cite{tidwell1992scaling,agnesi2002measurement}). The diffractive losses increase with increase in diode pump power and thus in our case it could be impossible to reach the threshold of the true CW Stokes generation by simply increasing the diode pump power. To obtain generation in the true CW mode one needs to lower the threshold to the values at which diffractive losses are small. In expression for the threshold (\ref{eqThresh}), one of the most crucial parameters is the losses product $\gamma _L \cdot \gamma_s$. The Stokes generation threshold as a function of the intracavity losses $\gamma_{int} = \gamma_{{int}_{L}}=\gamma_{{int}_{S}}$, is presented in Fig.~\ref{figThresh}. The results presented in Fig.~\ref{figThresh} are calculated with (\ref{eqThresh}) using the same assumptions and numerical values as utilized earlier for estimation of the output powers and efficiencies. For reference, results of the similar calculations for the 1.2~$\mu$m CWSSRL with the same parameters are also indicated in the Fig.~\ref{figThresh}.   When performing calculations for 1.2~$\mu$m CWSSRL we also assumed the quantities $w_L$  and $w_S$  were the same for the cases of 1.2~$\mu$m CWSSRL and 1.5~$\mu$m CWSSRL. These assumptions are not strictly correct because beam sizes in the cavity depend on the wavelength and on the heat load, which, in turn, depends on the diode pump power  and differs for lasing conditions at different transitions \cite{pavel2006neodymium}. However, it is a useful illustration to see that the threshold of a 1.5~$\mu$m CWSSRL is always higher than that of a 1.2~$\mu$m CWSSRL \textit{ceteris paribus}.

Fig.~\ref{figThresh} illustrates that the Stokes generation threshold grows rapidly with intracavity losses. The threshold of a 1.5~$\mu$m CWSSRL is always higher than that of a 1.2~$\mu$m CWSSRL and reaches high values for moderate intracavity losses. According to these calculations, for the intracavity losses $\gamma_{intr} = 3 \%$ (the value for the experimental laser described in Section~\ref{lExp}) the  threshold  reaches 13.2~W whereas for $\gamma_{intr} = 1 \%$  the threshold  is as low as 1.7~W.

In our view, the possibility of CW generation on $^4F_{3/2} \rightarrow ^4\!\!I_{13/2}$ transition in CWSSRL needs to be investigated further. A delicate balance between diode pump beam expansion (for minimizing aberrational thermal lensing) and optimal overlap of a pump beam with intracavity modes should be found. This may be accomplished through a special laser cavity design. The resonator should be designed in a way that maximizes overlap between a diode pump beam and intracavity laser and Stokes beams, and at least partially compensates for extremely strong thermal lensing in 1.5~$\mu$m CWSSRL. Thermal lensing could be managed by using a diode pump source operating at a longer wavelength (880~nm instead of 809~nm) to minimize quantum defect. Also, one can use laser host crystals with better thermal properties than those of the vanadates used in this work and/or optimize active ion concentration to decrease thermal effects \cite{pavel2006neodymium}. The generation threshold should be lowered by minimizing intracavity losses to $\gamma_{intra} < 1 \%$. This can be done by using active media of high optical quality and minimal impurities concentration and minimizing the number of separate optical elements inside the cavity. High-quality state-of-the-art optical mirrors with reflections at the fundamental wavelength $R_{1(2)}^{L}>99.99\mbox{x}\%$ should also minimize the threshold.



\section{Conclusion}
\label{lConclusion}

In conclusion, we demonstrated qCW (ms-long pulses) diode pumped solid state intracavity eye-safe Raman laser with fundamental radiation generated on $^4F_{3/2} \rightarrow ^4\!\!I_{13/2}$ transition of $\rm Nd^{3+}$ ions. The spectra of generation included two lines with the wavelengths of 1342~nm (fundamental line) and 1524~nm (Stokes line). The Stokes output power of 127~mW for a single direction corresponded to the maximum diode pump power of 56.8~W. Because both of the laser mirrors had identical spectral properties, we believe that identical power was generated in the opposite direction, and, thus, the overall output power at the Stokes wavelength reached 250~mW. The output power in qCW regime of operation could be scaled by using less reflective mirrors at the Stokes wavelength. The $\rm M^2$ parameters measured at 28~W of diode pump power were 2.7 and 6.5 for the Stokes and fundamental beams correspondingly. A true CW regime of operation is still challenging due to thermal issues but is possible in our view. Using a 880~nm diode pump, minimizing intracavity losses, and maximizing overlap between the interacting intracavity beams should improve thermal properties of the system and allow CW operation. The theoretical model of a CWSSRL developed in \cite{apanasevich2010power} was used to predict the threshold conditions of a 1.5~$\mu$m CWSSRLs and analyze output power characteristics of the 1.5~$\mu$m qCWSSRL developed. With intracavity losses set to 3~\%, the model showed satisfactory agreement with the experiment even with the constant beams radii used in calculations.

\section{Acknowledgements}
V.A.~Orlovich would like to acknowledge stimulating discussions with A.S.~Grabtchikov.
\label{lAcknowledgements}

\end{document}